\documentclass[useAMS,usenatbib,referee]{mn2e}
\usepackage{epsfig}
\usepackage{natbib}

\def \aj              {{\it Astron. J.}}

\def \aap             {{\it Astron. Astrophys.}}

\def \apj             {{\it Ap. J.}}
\def \apjl            {{\it Ap. J. Lett.}}

\def \apss            {{\it Astrophys. Space Sci.}}

\def \mnras           {{\it MNRAS}}
\def \nat             {{\it Nature}}

\title[Photometric detection of non-transiting low-mass companions]
{ Photometric detection of non-transiting short-period low-mass companions through the beaming, ellipsoidal and reflection effects in {\it Kepler} and CoRoT lightcurves}
\author[Faigler \& Mazeh]
{S.\ Faigler   and   T.\ Mazeh        
\\
School of Physics and Astronomy, 
Raymond and Beverly Sackler Faculty of Exact Sciences,\\ 
Tel Aviv University, Tel Aviv  69978, Israel
}

\date{submitted Feb 21st, 2011}

\pagerange{\pageref{firstpage}--\pageref{lastpage}} \pubyear{2007}

\def\LaTeX{L\kern-.36em\raise.3ex\hbox{a}\kern-.15em
    T\kern-.1667em\lower.7ex\hbox{E}\kern-.125emX}

\begin{document}
\label{firstpage}
\maketitle

\date{Received / Accepted }
\pagerange{\pageref{firstpage}--\pageref{lastpage}} \pubyear{2007}

\begin{abstract} 
We present a simple algorithm, BEER, to search for a combination of the BEaming, Ellipsoidal and the Reflection/heating periodic modulations, induced by short-period {\it non-transiting} low-mass companions.
The beaming effect is due to the increase (decrease) of the brightness of any light source approaching (receding from) the observer.
To first order, the beaming and the reflection/heating effects modulate the stellar brightness at the orbital period, with phases separated by a quarter of a period, whereas the 
ellipsoidal effect is modulated with the orbital first harmonic. The phase and harmonic differences between the three modulations allow the algorithm to search for a combination of the three effects and identify stellar candidates for low-mass companions.
The paper presents the algorithm, including an assignment of a likelihood factor to any possible detection, based on the expected ratio of the beaming and ellipsoidal effects,  given an order-of-magnitude estimate of the three effects.  
As predicted by Loeb \& Gaudi (2003) and Zucker, Mazeh \& Alexander (2007), the {\it Kepler} and the CoRoT lightcurves are precise enough to allow detection of massive planets and brown-dwarf/low-mass-stellar companions with orbital period up to 10--30 days. 
To demonstrate the feasibility of the algorithm, we bring two examples of 
candidates found in the first 33 days of the Q1 {\it Kepler} lightcurves. Although we used relatively short timespan, the lightcurves were precise enough to enable the detection of periodic effects with amplitudes as small as one part in $10^{4}$ of the stellar flux.

\end{abstract}


\section{Introduction}        
\label{introduction}                            

We present a simple algorithm, BEER, to search for a combination of the BEaming, Ellipsoidal and the Reflection/heating periodic modulations, induced by short-period {\it non-transiting} low-mass companions, using 
precise photometric stellar lightcurves. 
Two of the modulations are well known for many years from the study of close
binary stellar systems. These are the ellipsoidal variation \citep{morris85}, due to
the tidal distortion of each component by the gravity of its companion
\citep[see a review by][]{mazeh08}, and the reflection/heating variation
(referred here as the reflection modulation), induced by the
luminosity of each component that falls only on the close side of its
companion \citep[e.g.,][]{maxted02,harrison03,for10,reed10}.

A much smaller and less studied photometric modulation is the
 beaming effect, sometimes called Doppler boosting,
induced by the stellar radial motion.
This effect causes an increase (decrease) of the brightness of any light source approaching (receding from) the observer
 \citep[e.g.,][]{rybicki79}.
Before the era of space photometry this effect has been
noticed only once, by \citet{maxted00}, who observed
 KPD 1930+2752, a binary with a very short period, of
little longer than 2 hours, and a radial-velocity amplitude of 350 km/s. The
beaming effect of that system, which should be on the order of $10^{-3}$, was
hardly seen in the photometric data.

The beaming effect became relevant only recently, when 
space photometry, aimed to detect transits of exo-planets, has substantially
improved the precision of the produced lightcurves. 
The CoRoT \citep{rouan98, baglin06,auvergne09} 
and {\it Kepler} \citep{borucki10,koch10}  missions are producing
hundreds of thousands of continuous photometric lightcurves with
timespan of tens and hundreds of days, at relative precision level that can get
to $10^{-3}$--$10^{-4}$ per measurement, depending on the stellar brightness. 
It was therefore anticipated
that CoRoT and {\it Kepler} will detect each of the three modulations \citep[e.g.,][]{drake03,loeb03,zucker07}, for binaries and planets alike.

As predicted, already in the Q1 {\it Kepler} data, which spanned over only 33 days, \citet{vankerkwijk10} detected 
the ellipsoidal and the beaming effect of two eclipsing binaries, KOI-74 and KOI-81 \citep{rowe10}. 
\cite{carter10} detected the ellipsoidal, beaming and reflection effects in the {\it Kepler} lightcurve of the eclipsing binary KIC 10657664,
and derived the system parameters from the amplitudes of the three effects, determining the system to be comprised of a low-mass, thermally-bloated, hot white dwarf orbiting an A star. The effects were discovered even for brown-dwarf secondaries and planets.
\citet{welsh10} identified the ellipsoidal effect in the {\it Kepler} data of HAT-P-7, a system with a known planet of $1.8\,M_\mathrm{Jup}$ and a period of $2.2$
days \citep{pal08}. \citet{snellen09} detected in CoRoT data the 
reflection effect of CoRoT-1b.
\citet{mazeh10} detected the ellipsoidal and the beaming effect 
induced by CoRoT-3b, a $22\,M_\mathrm{Jup}$ massive-planet/brown-dwarf 
companion with a period of $4.3$ days \citep{deleuil08}.

However, space mission data can yield much more. In addition to eclipse events, CoRoT and {\it Kepler}
are producing data that can indicate the binarity of a system based on
the evidence coming only from the beaming, ellipsoidal and reflection
effects themselves. \citet{loeb03} suggested that the beaming
effect can be used to detect non-transiting exo-planets, and \citet{zucker07} extended 
this idea to binaries. \citet{loeb03} \citep[see also the discussion of][]{zucker07} showed
that for relatively long-period orbits, of the order of $10$--$100$ days,
the beaming modulation is stronger than the ellipsoidal and the
reflection effects, and therefore could be observed without the
interference of the other two modulations. However, the beaming
modulation by itself might not be enough to render a star a good
exo-planet candidate, as the pure sinusoidal modulation could be
produced by other effects, stellar modulations in particular \citep[e.g.,][]{aigrain04}.
The BEER detection algorithm, therefore, searches for stars that show in their space-obtained 
lightcurves some {\it combination} of the three modulations, the ellipsoidal and the
beaming effects in particular.

The CoRoT mission, and certainly the {\it Kepler}
satellite, have the required precision to reveal the ellipsoidal and
even the beaming modulations for massive planets and brown-dwarf/low-mass-stellar companions
with short enough orbital periods. This was demonstrated by \citet{mazeh10}
for the aforementioned CoRoT-3b, for which they detected the two modulations with amplitudes of about
60 and 30 ppm (parts per million), respectively.  

Searching for an unknown orbital period is more difficult than looking for
a modulation with known period and phase. However, 
the combination of at least two of the modulations, and their relative
amplitudes and {\it phases}, can suggest the presence of a small
non-transiting companion. Like in the transit searches, the candidates found 
must be followed by radial velocity (RV) observations, in order to confirm the existence
of the low-mass companion, and to reject the other possible interpretations of the photometric 
modulations.

This paper presents the details of the BEER algorithm. Section~2 presents
theoretical approximation of the beaming, ellipsoidal and reflection effects,
Section~3 explains the details and performance estimate of the algorithm itself, and Section~4 brings two 
candidates found in the {\it Kepler} Q1 data, with a companion mass of, up to the sine of the orbital inclination, $\sim$$70\,M_\mathrm{Jup}$. In a separate paper (Faigler, Mazeh et al., in preparation) we present RV observations that confirm the existence of the two companions.
Section~5 summarizes our results and argues that the BEER algorithm can discover short-period brown-dwarf companions and even massive planets, given the CoRoT and {\it Kepler} data accuracy.

\section{Theoretical approximation of the three effects}        
To perform the search for massive planets and brown-dwarf/low-mass-stellar companions we
need order-of-magnitude approximations for the three effects. For the BEER
algorithm, we use the expressions listed by \citet{mazeh10}
for circular orbits, assuming the companion is much smaller than the primary star, and therefore ignoring its luminosity.
The expressions for a $10\,M_\mathrm{Jup}$ companion become
\begin{equation}
A_\mathrm{beam} =
\alpha_\mathrm{beam}\, 4\frac{K_{\scriptscriptstyle\mathrm RV}}{c} 
=
27 \ \alpha_\mathrm{beam}
\left(\frac{M_*}{M_{\odot}}\right)^{-2/3}
\left(\frac{P_\mathrm{orb}}{1 \,{\rm day}}\right) ^{-1/3}
\left(\frac{m_2 \sin i}{10\,M_\mathrm{Jup}}\right) \ {\rm ppm},
\end{equation}

\begin{equation} 
 A_\mathrm{ellip}  =
\alpha_\mathrm{ellip}
\frac{m_2 \sin i}{M_*} \left(\frac{R_*}{a}\right)^3 \sin i
=
128 \ \alpha_\mathrm{ellip} \sin i
\left(\frac{R_*}{R_{\odot}}\right)^3
\left(\frac{M_*}{M_{\odot}}\right)^{-2}
\left(\frac{P_\mathrm{orb}}{1 \ {\rm day}}\right) ^{-2}
\left(\frac{m_2 \sin i}{10\,M_\mathrm{Jup}}\right) \ {\rm ppm},
\end{equation} 


\begin{equation}
A_\mathrm{refl} =
\alpha_\mathrm{refl}\, 0.1
\left(\frac{r_2}{a} \right)^2 \sin i
=
57 \ 
\alpha_\mathrm{refl} \sin i
\left(\frac{M_*}{M_{\odot}}\right)^{-2/3}
\left(\frac{P_\mathrm{orb}}{1 \, {\rm day}}\right) ^{-4/3}
\left(\frac{r_2}{R_\mathrm{Jup}}\right)^2  \ {\rm ppm}.
\end{equation}
%
In the above expressions, $m_2$ and $r_2$ are the companion mass and radius, 
$M_*$ and $R_*$ are the primary mass and
radius, $P_\mathrm{orb}$ and $a$ are the orbital period and
semi-major axis, $K_{\scriptscriptstyle\mathrm RV}$ 
is the semi-amplitude of the stellar
radial-velocity modulation induced by the companion, $i$ is the
orbital inclination relative to our line of sight, and $c$ is the
speed of light.

The $\alpha$'s represent order-of-unity coefficients that necessitate
more detailed model to evaluate:

\begin{itemize}

\item
The expression for the beaming effect includes two factors. The
$4K_{\scriptscriptstyle\mathrm RV}/c$ 
factor represents the beaming effect for bolometric
photometric observations, but ignores the Doppler shift photometric
effect, which appears when the photometric observations are made in a
specific bandpass, so that some of the stellar light is shifted out of
or into the observed bandpass. The latter is accounted for by the
$\alpha_\mathrm{beam}$ factor. We assume here black body stellar radiation model which
yields for the CoRoT and the {\it Kepler} bandpasses and for F-G-K spectral-type stars,  $\alpha_\mathrm{beam}$  value between $0.8$ and $1.2$. 

\item
The $\alpha_{\mathrm ellip}$ factor represents the response of the
stellar surface to the tidal effect induced by the companion, and to
first order can be written \citep{morris85} as

\begin{equation} 
 \alpha_\mathrm{ellip}
\simeq
0.15\frac{(15+u)(1+g)}{3-u},
\end{equation}
where $g$ is the stellar gravity darkening coefficient, whose expected range is $0.3$--$1.0$, and $u$ is the limb-darkening coefficient, whose range is $0$--$1$ and is typically
$\approx 0.6$ for solar-like stars \citep[see, for example,][]{mazeh08}. Thus, we
estimate, that for F-G-K spectral-type stars, $\alpha_\mathrm{ellip}$ value is between $1.0$ and $2.4$. 

Note that the expression for $A_\mathrm{ellip}$ includes an extra $\sin i$
factor, in addition to the one that appears in the $m_2 \sin
i$ factor of Equation~2. This reflects the stronger dependence of the
ellipsoidal modulation on the inclination angle. If we know well
enough all the other factors of $\alpha_\mathrm{ellip}$ and
$\alpha_\mathrm{beam}$ and we derive $A_\mathrm{ellip}$ and
$A_\mathrm{beam}$, we can, at least in principle, 
estimate both $m_2 \sin i$ {\it and} $\sin i$.

\item
In our simplistic approximation we include in the reflection modulation 
the thermal emission from the dayside of the companion, 
assuming both are modulated with the same phase \citep[e.g.,][]{snellen09}. 
This approximation does not model the small phase shift of the reflection modulation 
that can be present if the companion is not tidally locked, or if there is advection of
heat away from the sub-stellar point, as shown by \citet{knutson07}.
The amplitude of the modulation of the reflected light alone is 
\begin{equation}
A_\mathrm{refl}
=
p_\mathrm{geo}
\left(\frac{r_2}{a} \right)^2 \sin i,
\end{equation}
where $p_\mathrm{geo}$ is the geometrical albedo. \citet{rowe08}
found quite small albedo, of $0.03$, for HD~209458b, but recent study \citep{cowan11}
 suggested that exo-planets could have much larger
albedo, up to $0.5$. We therefore estimate, somewhat arbitrarily, the
geometrical albedo coefficient by
$p_\mathrm{geo}=0.1\alpha_\mathrm{refl}$, and estimate that the value
of $\alpha_\mathrm{refl}$ can be between $0.2$ and $5$.  

\end{itemize}

\section{The BEER algorithm}        
\label{algorithm}                            

\subsection{Two-harmonic search}

Before searching for small periodic effects in any stellar lightcurve, 
we have to prepare and clean the data. This is done in two stages. In the
first one we remove jumps and outliers, and in the second stage we identify and subtract the long-term
variation of the lightcurve with the discrete cosine
transform \citep{ahmed74}, adopted to unevenly spaced data. 
 This is done by subtracting from the data a linear model of all cosine functions
with frequencies from zero to $0.1$ d$^{-1}$, with frequency separation of $1/(2T)$, $T$ being the 
total time span of  the lightcurve \citep[see details of our approach in][]{mazeh10}.  

We then proceed to fit the data with a model that includes the
ellipsoidal, beaming and reflection effects for every possible period,
$P_\mathrm{orb}$. The algorithm approximates each of the three effects
by pure sine/cosine function, relative to phase zero taken at the time
of conjunction --- $t_\mathrm{conj}$, when the small companion is in
front of the stellar component of the system. This fiducial point
replaces the time of transit --- $t_\mathrm{tran}$, used when modeling
transiting planets and eclipsing binaries \citep[e.g.,][]{mazeh10}.
The reflection and the beaming effects are approximated by
cosine and sine functions, respectively, with the orbital period, and
the ellipsoidal effect by a cosine function with half the orbital
period. In this approximation we express the stellar flux
modulation $\Delta F$ as a fraction of the averaged flux ${\bar{F}}$,
and as a function of $\hat{t}\equiv t-t_\mathrm{conj}$:

\begin{equation} 
\frac{\Delta F_\mathrm{ellip}(\hat{t})}   {\bar{F}} =
-A_\mathrm{ellip}\cos\left(\frac{2\pi}{P_\mathrm{orb}/2}\hat{t}\right),
\end{equation} 
  
\begin{equation} 
\frac{\Delta F_{\mathrm{beam}}(\hat{t})}            {\bar{F}}  =
A_\mathrm{beam}\sin\left(\frac{2\pi}{P_\mathrm{orb}}\hat{t}\right),
\end{equation} 
  
\begin{equation} 
\frac{\Delta F_{\mathrm{refl}}(\hat{t})}{\bar{F}} =
-A_\mathrm{refl}\cos\left(\frac{2\pi}{P_\mathrm{orb}}\hat{t}\right),
\end{equation} 
where the coefficients, $A_\mathrm{ellip}$, $A_\mathrm{beam}$ and
$A_\mathrm{refl}$ are all positive.

We note that contrary to the case of transiting planets, like CoRoT-3b,
we do {\it not} know {\it a priori} the time of $t_{conj}$.  
The algorithm therefore
fits the cleaned, detrended lightcurve with a linear five-parameter
model, consisting of two frequencies:

\begin{equation}
\mathcal{M}(\bar{t})=a_0
+a_\mathrm{1c}\cos\left(\frac{2\pi}{P_\mathrm{orb}}\bar{t}\right)
+a_\mathrm{1s}\sin\left(\frac{2\pi}{P_\mathrm{orb}}\bar{t}\right)
+a_\mathrm{2c} \cos\left(\frac{2\pi}{P_\mathrm{orb}/2}\bar{t}\right)
+a_\mathrm{2s}\sin\left(\frac{2\pi}{P_\mathrm{orb}/2}\bar{t}\right),
\end{equation}
where $\bar{t}$ is the time relative to some arbitrary zero time. Since in this model
the amplitudes are free to be either positive or negative, the usual bias of over-estimating a sinusoid's amplitude when fitting noisy data is not present.
 
As explained, we expect $a_\mathrm{2s}$ to be close to zero and $a_\mathrm{2c}$ to be negative relative to the time of conjunction, so after the
fitting is done we find $t_\mathrm{conj}$ that results in $a_\mathrm{2s}=0$ and $a_\mathrm{2c}$ being negative. This time is when the first harmonic model 
\begin{equation}
\mathcal{M}_2(\bar{t})=
a_\mathrm{2c} \cos\left(\frac{2\pi}{P_\mathrm{orb}/2}\bar{t}\right)
+a_\mathrm{2s}\sin\left(\frac{2\pi}{P_\mathrm{orb}/2}\bar{t}\right)
\end{equation}
has its minimum. We note that because $\mathcal{M}_2$ presents the first harmonic component of the model, it has two minima per period. The algorithm choice between the two minima is detailed below.

The algorithm then performs a new linear fit with a four-parameter model:

\begin{equation}
\mathcal{M}(\hat{t})=
\hat{a}_0
+\hat{a}_\mathrm{1c}\cos\left(\frac{2\pi}{P_\mathrm{orb}}\hat{t}\right)
+\hat{a}_\mathrm{1s}\sin\left(\frac{2\pi}{P_\mathrm{orb}}\hat{t}\right)
+\hat{a}_\mathrm{2c} \cos\left(\frac{2\pi}{P_\mathrm{orb}/2}\hat{t}\right),
\end{equation}
which is our final model for that period.

\subsection{A likelihood parameter}

So far, the search performs a regular double-harmonic search \citep[e.g.,][]{shporer06}, and therefore the fitting process could find any values for the three amplitudes. We now  exercise our astrophysical expectations for the amplitudes and phases of the three effects, and assign to each period a likelihood factor which expresses how likely are the derived ratio between the amplitudes of the beaming and the ellipsoidal effects, and the phase difference between the beaming and the reflection effects. 

 If indeed the modulation, with its double harmonic components, is induced by a low-mass companion with negligible luminosity, we expect $\hat{a}_\mathrm{1s}$ to
represent the beaming effect and therefore to be positive and
 $\hat{a}_\mathrm{1c}$ to represent the reflection effect
and therefore be negative.  The $\hat{a}_\mathrm{1c}$ coefficient is negative by our definition of $t_{conj}$.  

The algorithm therefore distinguishes between two cases:
\begin{itemize}
\item
The beaming and the reflection coefficients, $\hat{a}_\mathrm{1s}$  and $\hat{a}_\mathrm{1c}$, respectively, have opposite signs. In this case 
the algorithm chooses (between the two options, see above) $t_{conj}$  so that 
$\hat{a}_\mathrm{1s}$ is positive and 
 $\hat{a}_\mathrm{1c}$ is negative.

\item
The beaming and the reflection coefficients, $\hat{a}_\mathrm{1s}$  and $\hat{a}_\mathrm{1c}$, respectively, have the same sign. In such a case the algorithm chooses 
(between the two options, see above) $t_{conj}$ such that the more significant coefficient has the correct sign. The other coefficient is then set to zero.
\end{itemize} 

Our model is therefore composed of two or three components, depending on the relative signs of the sine and cosine components of the fitting. Because we are looking for a system that displays {\it both} beaming and ellipsoidal modulations, we consider as our 
model goodness-of-fit parameter the r.m.s. of the smaller between these two fitted modulations. To scale the goodness-of-fit we divide the model r.m.s. by that of the residuals relative of the total model. This definition implies that a prominent peak in the periodogram indicates that  {\it both} modulations, the ellipsoidal {\it and} the beaming ones, are significant at the specific peak period.  
This ratio, derived for every possible period, marks the first stage of our periodogram. 

We now proceed to assign a likelihood factor to the model. 
Using Equations (1) \& (2) we get for the ratio between the amplitudes of the ellipsoidal and the beaming effects: 

\begin{equation}
\mathcal{R} {\equiv} \frac{A_{ellip}} {A_{beam}}  
=5
\frac{\alpha_{ellip}}  {\alpha_{beam}} 
\left(\frac{M_*}{M_\odot}\right)^{-4/3}
\left(\frac{R_*}{R_\odot}\right)^{3} 
\left(\frac{P_{orb}}{1\, \rm day}\right)^{-5/3} 
{\sin i} .
\end{equation}
We note that the amplitude ratio of the two effects depends only on parameters associated with the stellar properties of the primary star and the orbital period and inclination, and does not depend on the companion mass. As such, this ratio can serve as basis for comparing and validating the relevance
of the detected amplitudes. 

To do that, we distinguish between the {\it expected} ratio

\begin{equation}
\mathcal{R}_\mathrm{exp}
{\equiv} 5
\frac{\alpha_{ellip}}  {\alpha_{beam}} 
\left(\frac{M_*}{M_\odot}\right)^{-4/3}
\left(\frac{R_*}{R_\odot}\right)^{3} 
\left(\frac{P_{orb}}{1\, \rm day}\right)^{-5/3} 
{\sin i}, 
\end{equation}
and the {\it observed} ratio,
$\mathcal{R}_\mathrm{obs}{\equiv}|\hat{a}_{\mathrm{2c}}|/|\hat{a}_{\mathrm{1s}}|$.
Obviously, $\mathcal{R}_\mathrm{obs}$ is not known exactly, but can be described by a probability distribution, 
 $Pr_{\mathrm obs}\left(\mathcal{R}\right)$, which 
depends on the precision of the two derived amplitudes.

Furthermore, our expectation for $\mathcal{R}_\mathrm{exp}$ also does not have a single value, due to lack of precise knowledge of its factors. However, we can generate for every star and period
a likelihood function 
 $\mathcal{F}_{\mathrm exp}\left(\mathcal{R}\right)$, 
which assigns a likelihood value to a range of possible values of $\mathcal{R}$. This function reflects the random distribution of the inclination, together with our prior distributions of the mass and radius of the given star, and the prior likelihood, depending on the theory, of a range of values of $\alpha_{beam}$ and $\alpha_{ellip}$. The function 
$\mathcal{F}_{\mathrm exp}\left(\mathcal{R}\right)$ 
 thus, in essence, encompasses our expectations of the system, and serves as a prior, defined such that its maximum is unity and minimum is zero.

To demonstrate the situation we plot the two functions in Figure~1
for the {\it Kepler} star 
K06521917, derived for a period of 1.345 day. 
In this example, the most probable $\mathcal{R_\mathrm{exp}}$ value is $10.01$, while $\mathcal{R}_\mathrm{obs}=1.4\pm0.14$. The likelihood in this case is quite small --- $0.07$.

\begin{figure*} 
\centering
\resizebox{12cm}{9cm}
{
\includegraphics{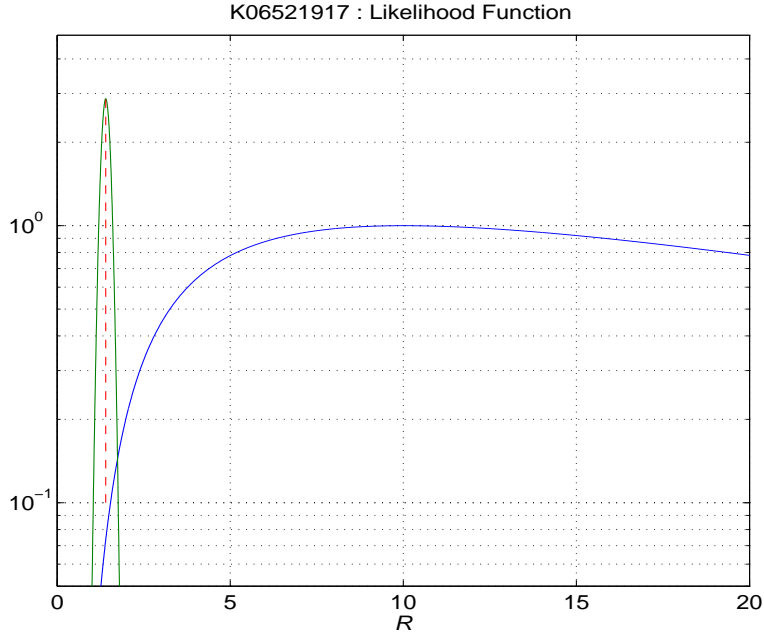}
}
\caption{
 The likelihood function of the expected value of $\mathcal{R}_\mathrm{exp}$ (blue) and the probability distribution of the observed value of $\mathcal{R}_\mathrm{obs}$ (green) for the {\it Kepler} star 
K06521917, derived for a period of 1.345 day.  
The likelihood function peaks at $\mathcal{R}=10.01$,  and the 
probability distribution peaks at $\mathcal{R}=1.4$, denoted by a vertical dashed line.
}
\end{figure*}

For a given period, the value of the likelihood factor, defined as

\begin{equation}
\mathcal{L}\left(P\right)=
\int{\mathcal{F}_{\mathrm exp}\left(\mathcal{R}\right)
\times Pr_{\mathrm obs}(\mathcal{R})
d\mathcal{R}} ,
\end{equation}
determines how likely are the two modulations to be caused by a low-mass companion for any given period. The likelihood factor is derived by integration over the probability distribution of  $\mathcal{R}_\mathrm{obs}$, weighted by the likelihood function.   
The final periodogram is the  goodness-of-fit of the
model multiplied by the likelihood factor $\mathcal{L}\left(P\right)$, generated for range of periods. The highest peak of each periodogram is our best estimate for the orbital period of the presumed low-mass companion.

\subsection{Significance and detection limit}

For any derived periodogram with its highest peak, we have to answer two questions:

\begin{itemize}
\item
Is this peak significant, representing a real periodic modulation, or is it a result of random noise?
\item
If the peak is real and the stellar lightcurve does contain a periodic modulation, is this modulation induced by a low-mass companion?
\end{itemize} 

The significance of a period detection can be estimated, for example, by the ratio of the highest peak to the second highest one in the periodogram, not including the harmonics of the highest peak. We choose to put our threshold detection when this ratio is 2. Bootstrap simulations of  10,118 {\it Kepler} Q1 stars with magnitude within $12$--$13$ range, and with radius smaller than $3R_{\odot}$,  did not yield a single false detection using the factor 2 threshold, indicating a 99.99\% significance.   
If a peak above this threshold  is derived, we consider the corresponding star as a {\it candidate} host of low-mass companion, with the orbital period corresponding to the peak frequency.

Unfortunately, answering the second question is more difficult. Even with a highly significant peak at the periodogram, at this point we can not rule out false positive detections, which can rise from stellar modulations of some kind. Therefore, in order to confirm the detection of a low-mass companion, we do need follow-up RV observations. As the presumed period is known, a few measurements should be enough to confirm or reject the low-mass conjecture.

We now turn to roughly estimate our detection rate for the bright-star lightcurves of {\it Kepler}.
 To do that we took the actual Q1 lightcurves and added to them  
simulated beaming, ellipsoidal and reflection effects of binaries with a period of $3.2$ days and randomly chosen phases.
In this way, we used the real noise characteristics of the {\it Kepler} lightcurves to estimate the detection performance of the algorithm. The sample was composed of the 10,118 {\it Kepler} Q1 stars with magnitude within $12$--$13$ range, and with radius smaller than $3R_{\odot}$, and the simulated effects were prepared with $\alpha_\mathrm{beam}=1$, $\alpha_\mathrm{ellip}=1.5$,  $\alpha_\mathrm{refl}=1$, and $\sin i=\pi/4$, using the stellar masses and radii available from the {\it Kepler} catalog, all with the same secondary mass. 
 We repeated the simulation for six different secondary masses --- $10/20/40/80/160/320 \, M_\mathrm{Jup}$. In each simulation we applied the BEER algorithm
and counted the stars for which the highest periodogram peak was at least twice as much higher than the next one, and we could detect both the beaming and the ellipsoidal modulations with $5\sigma$ significance.
If the highest peak in our periodogram did not have the inserted period we did not consider that star as a detection.

\begin{figure*} 
\centering
\resizebox{12cm}{9cm}  
{ \includegraphics{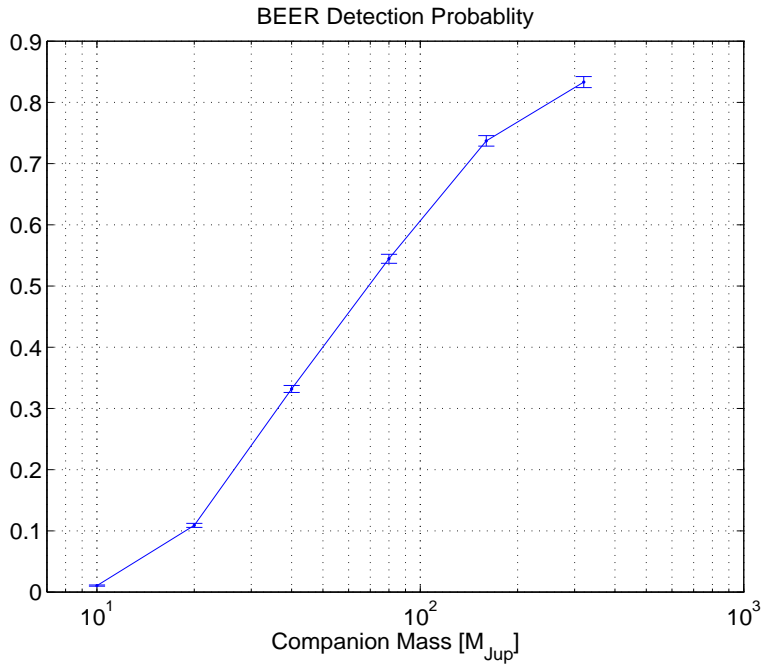}}
\caption{
 Percentage  of $5\sigma$ detections of the BEER algorithm in our simulations, as a function of
the secondary mass. Each of the simulations included a set of 10,118 {\it Kepler} Q1 stars, with magnitude within $12$--$13$ range, and with radius smaller then $3R_{\odot}$ (see text).
}
\end{figure*}

The results are given in  
Figure~2, which indicates the fraction of systems that have been detected in the simulations as a function of the mass of the inserted modulation. We note that while the algorithm detected only one percent of the simulated massive planets, with masses of 
$10 M_{\mathrm Jup}$, almost all simulated modulations corresponding to a secondary mass of 
$\sim 0.3 M_{\odot}$ were detected. The detection rate derived in this way indicates the potential of the BEER algorithm. We note that some of the non-detections may be caused by systems with {\it real} companions with different periods, with masses higher than the ones inserted by the simulation. This is specially true for the simulated $10 M_{\mathrm Jup}$ cases.

\section{Two simple examples}        
\label{examples}                            

In order to demonstrate the effectiveness of the BEER algorithm, this section presents
the detection of both the beaming and ellipsoidal modulations in two different {\it Kepler} lightcurves, induced by two low-mass companions. 
The two stars, KIC 08016222 (hereafter K6222) and KIC 010848064 (hereafter K8064), were found by the BEER algorithm, applied to the $11,249$ brightest stars in the {\it Kepler} public lightcurves database (http://archive.stsci.edu/kepler/),
 with available stellar mass and radius estimate. In both lightcurves BEER detected periodic modulations, which we attributed in both cases to variations induced by a low-mass companion estimated, up to $\sin i$, at $\sim$$70\,M_\mathrm{Jup}$. 
We chose to bring these two examples to demonstrate the precision of the {\it Kepler} mission, which allows detecting with high significance a companion with mass that could be in the lower end of the stellar mass range.
In a separate paper (Faigler, Mazeh et al., in preparation) we present RV observations that confirm the existence of the two companions.

Figure~3 shows the obtained {\it Kepler} flux variation of the two stars, each of which is divided by its own mean flux. The lightcurves include only the Q1 released data, which lasted for 33 days, from May 13, 2009 until June 15, 2009.

\begin{figure*} 
\centering
\resizebox{16cm}{8cm}
{
\includegraphics{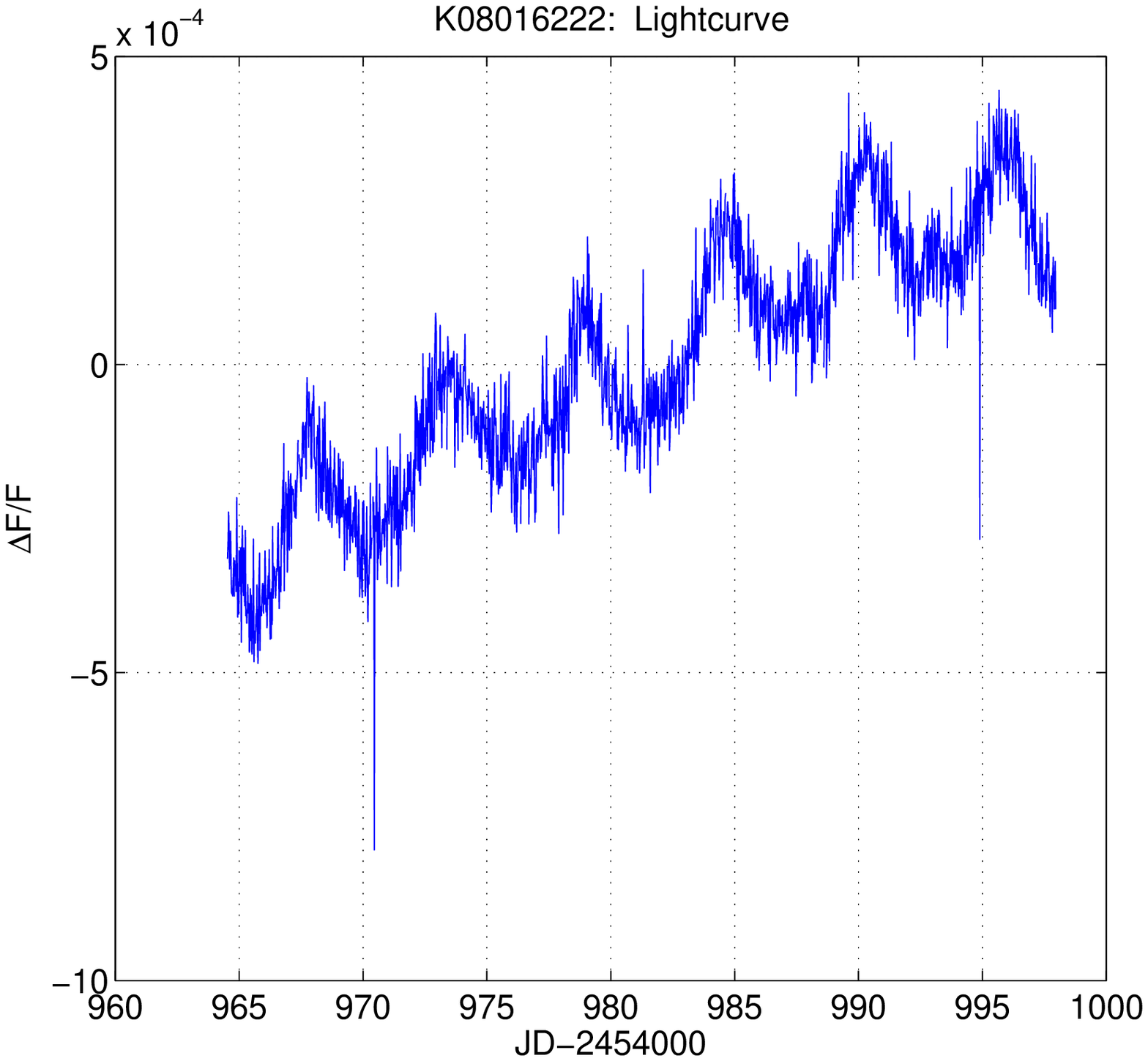}
\includegraphics{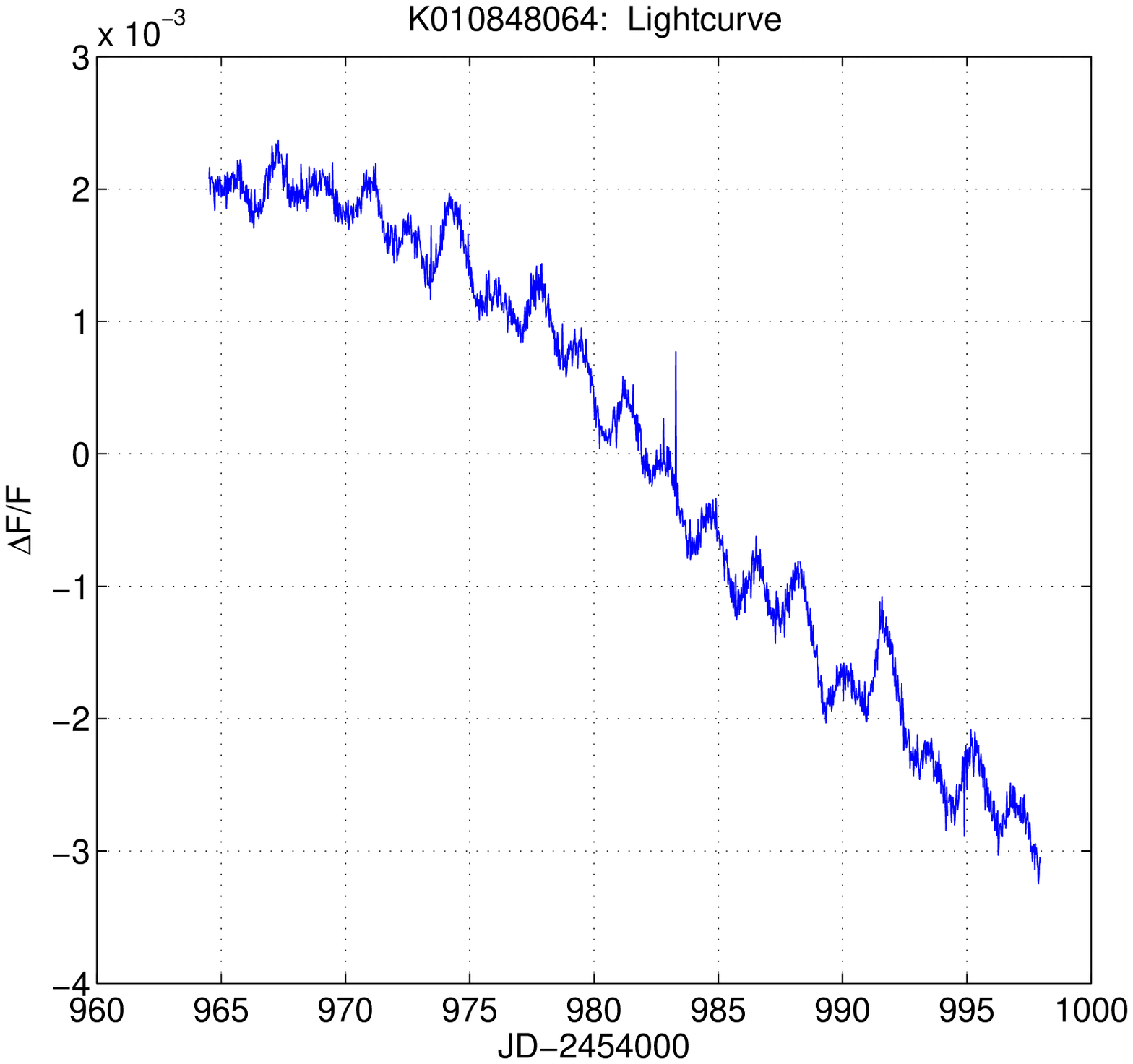}
}
\caption{
The original Q1 {\it Kepler} lightcurves of K6222 and K8064. The flux variation of each star is divided by its own mean.
Left: K6222. Right: K8064. (Note the scale difference of the two panels).
}
\end{figure*}

In both lightcurves one can see a clear periodic modulation superposed on a long-term variability of the stellar flux. Figure~4 shows the BEER periodograms of the two stars, using the cleaned detrended lightcurves (see Section 3.1). Both periodograms show a prominent peak, indicating the presence of a low-mass companion. 

 Figure~5 presents the likelihood function of the two stars for the best period found by the periodogram. 
The likelihood factor, based on Equation (14), is $0.99$ and $0.97$ for K8064 and K6222, respectively, implying that the detected periodic modulation {\it could} have been induced by a low-mass companion.
Figure~6 presents the folded cleaned lightcurves, with phase zero defined to coincide with $t_\mathrm{conj}$ found by our algorithm for each star, together with the fitted BEER model. 

\begin{figure*} 
\centering
\resizebox{16cm}{8cm}
{
\includegraphics{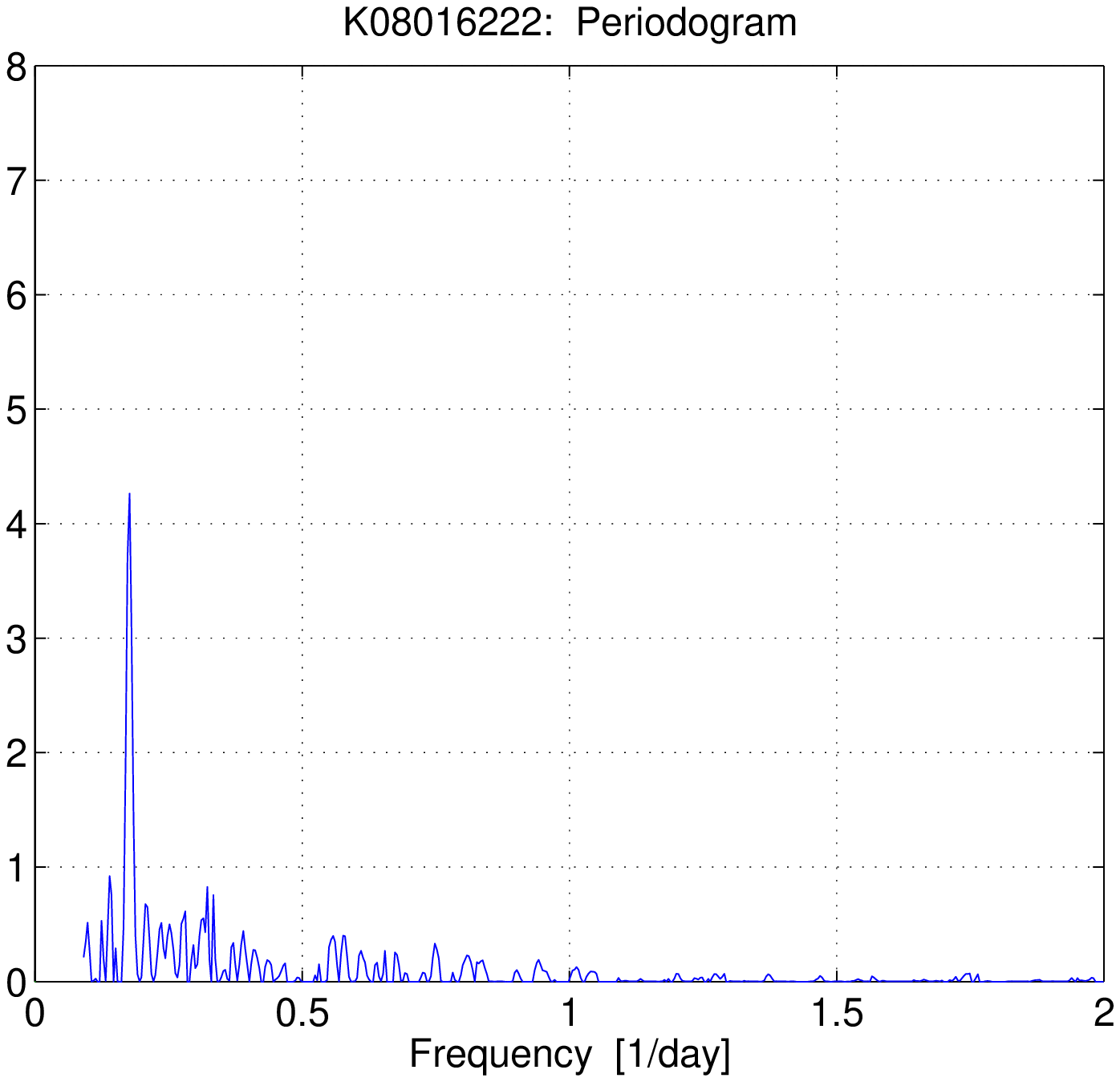}
\includegraphics{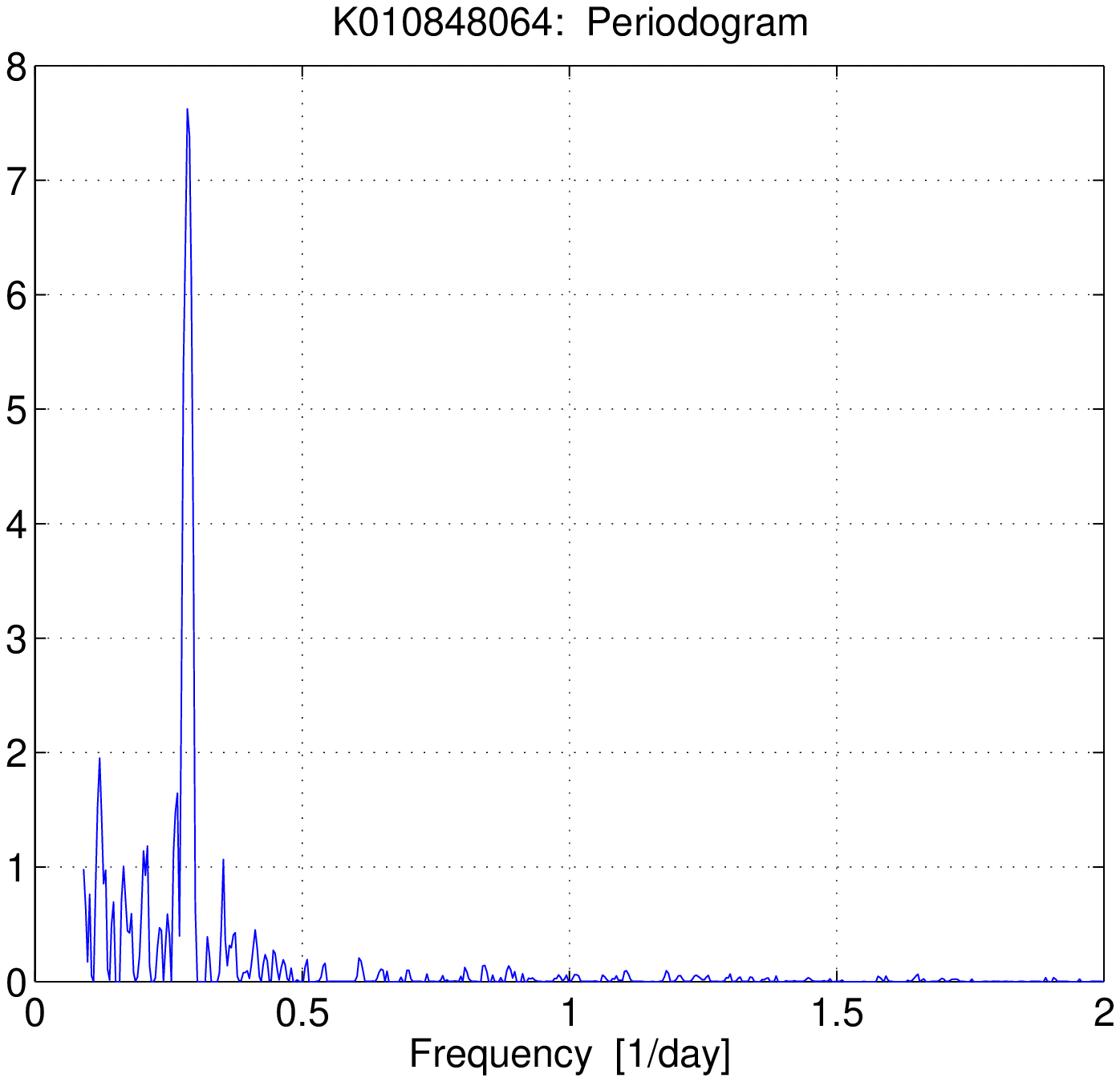}
}
\caption{
The derived periodograms of the BEER algorithm for the cleaned detrended lightcurves. 
Left: K6222. Right: K8064.
}
\end{figure*}

\begin{figure*} 
\centering
\resizebox{16cm}{8cm}
{
\includegraphics{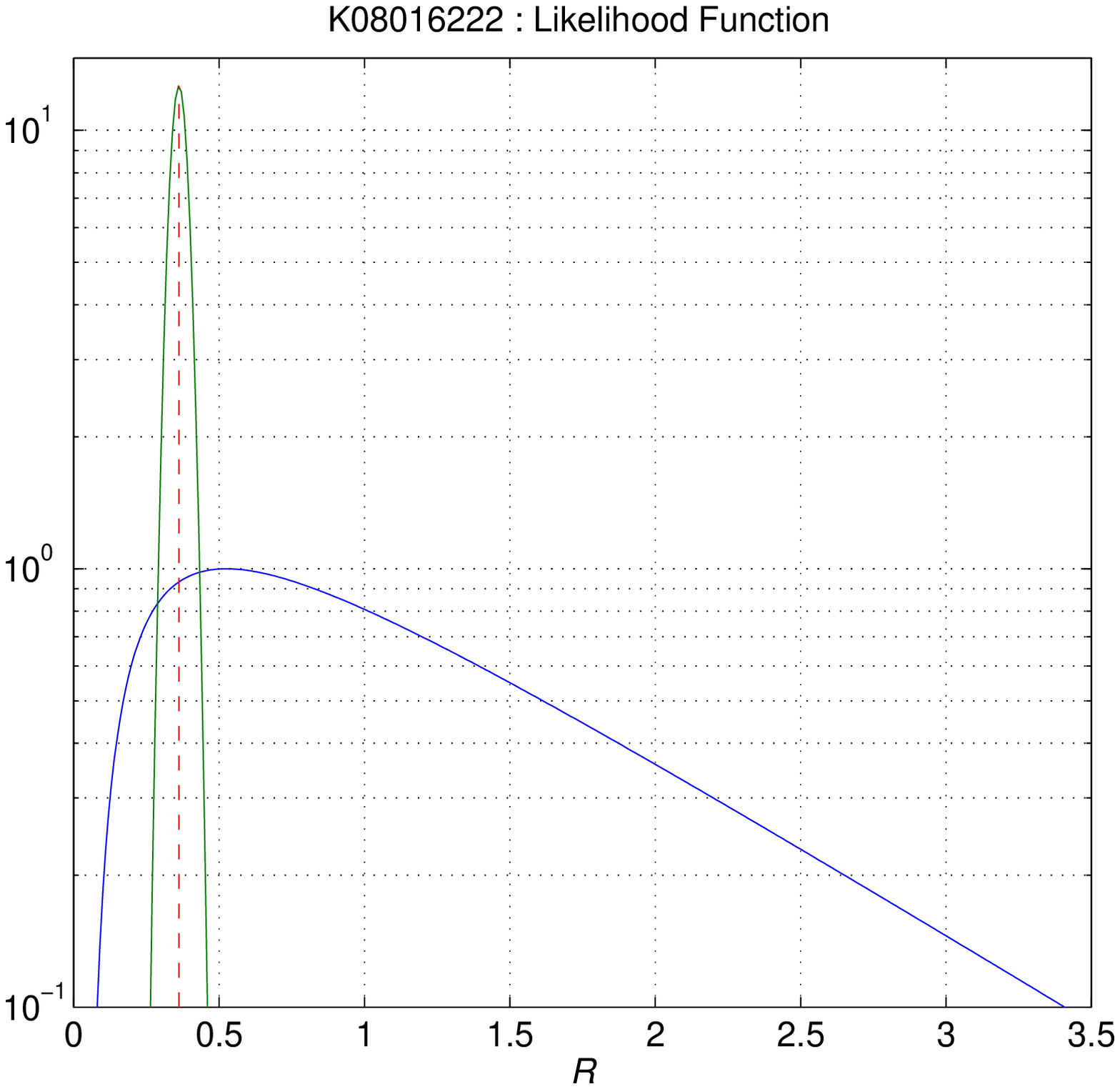}
\includegraphics{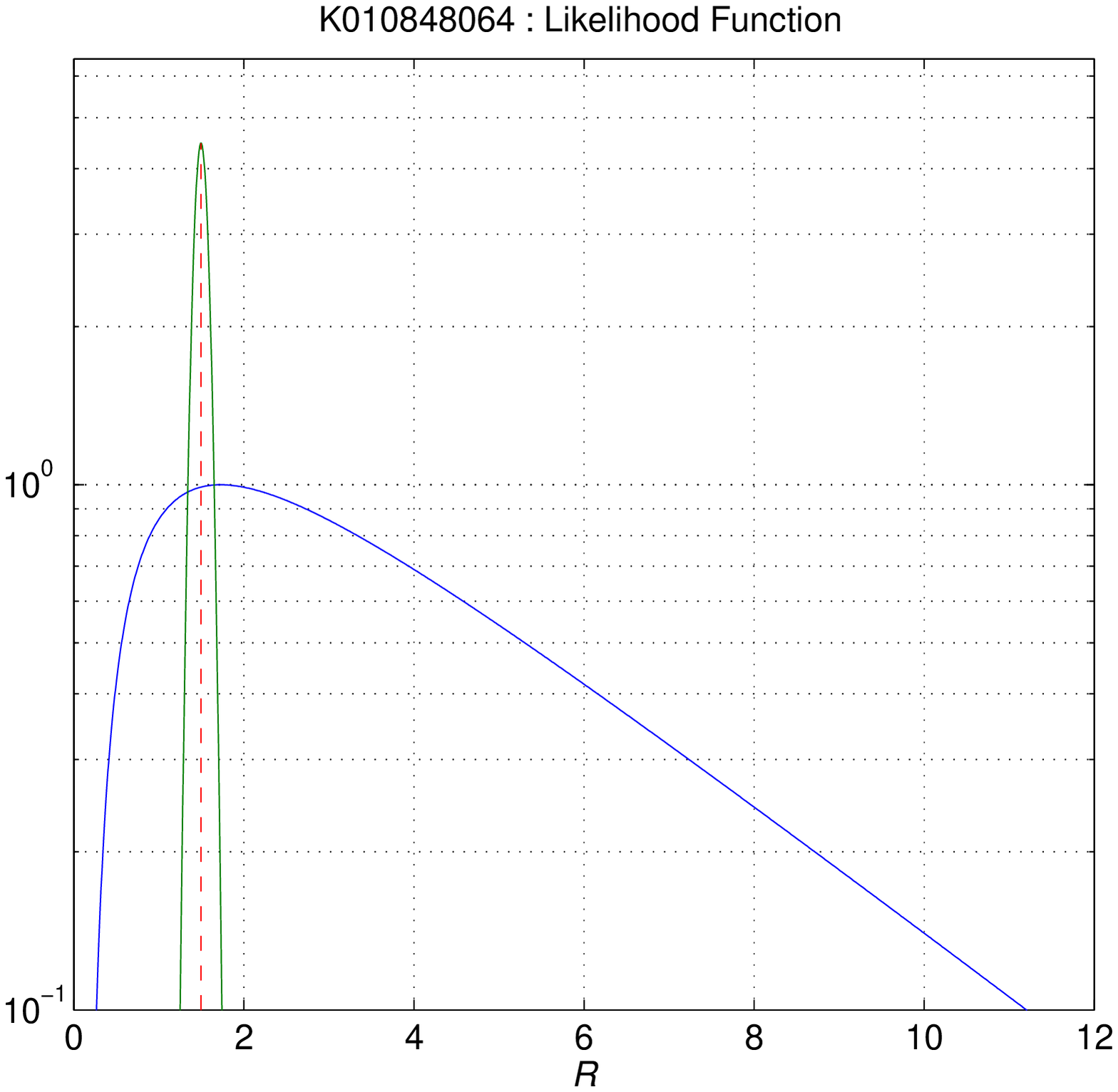}
}
\caption{
 The likelihood function of the expected value of $\mathcal{R}_\mathrm{exp}$ (blue) and the probability distribution of the observed value of $\mathcal{R}_\mathrm{obs}$ (green), for the two examples. 
Left: K6222. Right: K8064. (Note the scale difference of the two panels).
}
\end{figure*}

\begin{figure*} 
\centering
\resizebox{16cm}{8cm}
{
\includegraphics{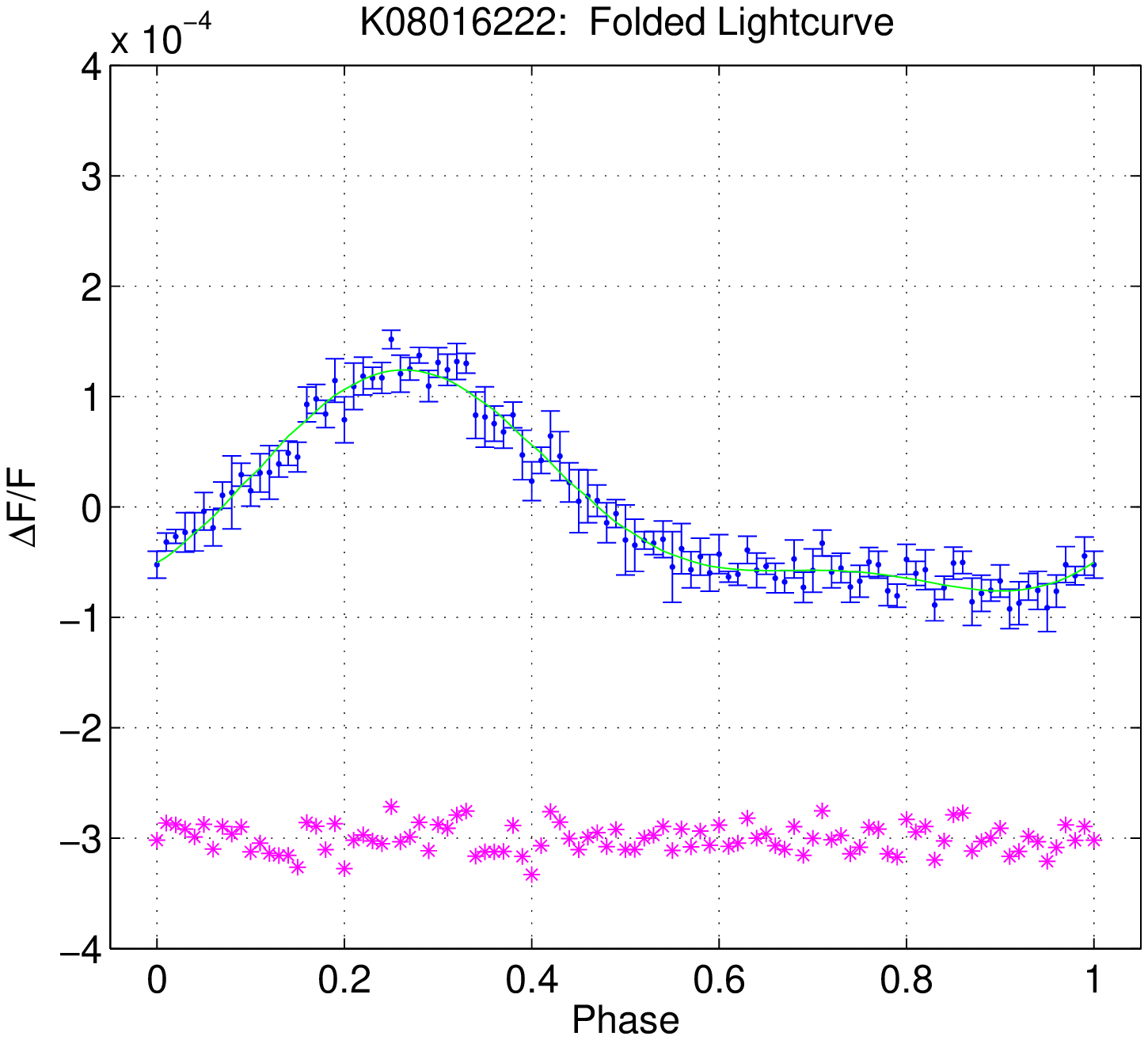}  
\includegraphics{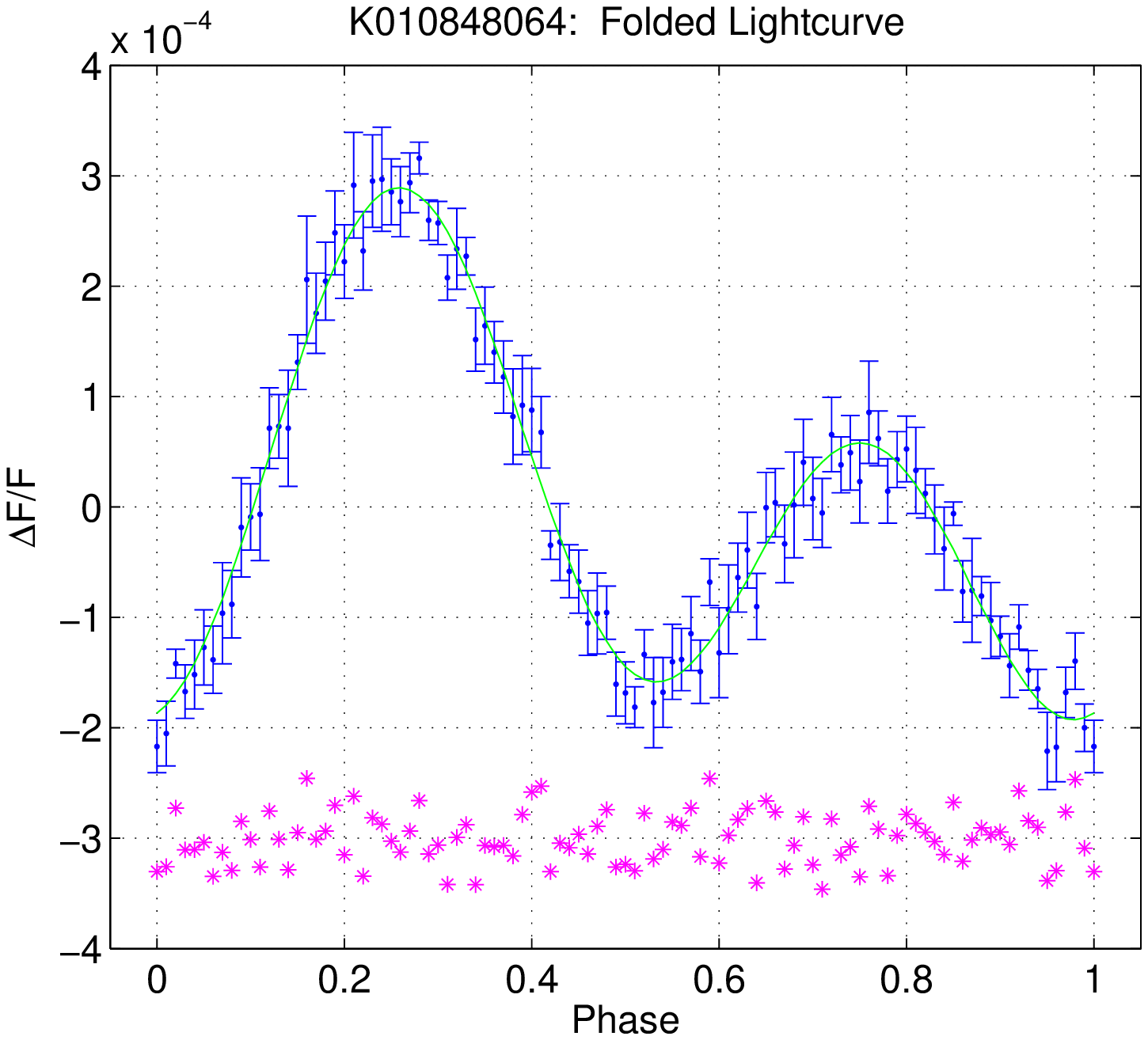}
}
\caption{
The folded cleaned lightcurves, binned into 100
bins, with the fitted model. The errors of each bin represent $1\sigma$ scatter of the measurements in each bin. The continuous line presents the BEER model. The residuals from this model are plotted at the bottom of
the figure.  Both panels have the same scale.
Left: K6222. Right: K8064.
}
\end{figure*}

Table~1 brings some details of the primaries of the two stars --- estimated mass and radius and their {\it Kepler} magnitudes (http://archive.stsci.edu/kepler/), and presents the periods and derived amplitudes of the three effects found by the BEER algorithm for the two systems. It also brings the derived masses of the unseen companions, and the expected amplitude of the RV modulation.

\begin{table}
\caption{
The derived parameters of the three effects for K6222 \& K8064 }

\begin{tabular}{lrcl}
                 &          K6222   & K8064 &  \\
\hline
$M_*$        & $1.1$                  & $1.2$      &   $ M_{\odot}$\\
$R_*$         & $1.3$                  & $1.5$       &   $ R_{\odot}$\\
Kp             &  $11.6$                &  $12.1$    & mag  \\
\hline
Period        &$5.6 \pm 0.2$   &  $3.53 \pm 0.07$ & days\\ 
Ellipsoidal   &$34\pm 2$          &   $170\pm5$       & ppm\\
Beaming    &$96\pm 2$          &   $116\pm5$         &  ppm  \\
Reflection  &$11\pm 2$           &    $17\pm5$         & ppm \\
 \hline
$m_2 \sin i$         &$\sim$$70$       &   $\sim$$70$ &   $ M_{\mathrm Jup}$\\
Expected $K_{\mathrm\scriptscriptstyle RV}$ & $7$    &     $9$  & km/s                       \\
\hline
\end{tabular}
\label{table_coeff}
\end{table}

One can see from Table~1 that the relative strength of the beaming and the ellipsoidal effects is quite different in the two stars. While for K8064 the derived amplitude of the ellipsoidal effect is larger than that of the beaming modulation, in the case of K6222, the derived beaming effect is almost three times larger than the ellipsoidal modulation. This difference changes the appearance of the folded lightcurves. This is so because 
the ellipsoidal effect is symmetric around phase $0.5$, while the beaming effect is anti-symmetric. Therefore, the shape of the folded lightcurve of K8064 looks almost symmetric, while in the case of K6222  the symmetric appearance is completely lost. 

The different ratio of the two amplitudes is rooted in Equation 12, which shows that this ratio depends on the stellar parameters of the primary, on the stellar radius to the third power in particular, and on the orbital period. { As the {\it Kepler} estimate of the radius of K8064 is larger than that of K6222 and the derived orbital period is shorter, we expect the amplitude ratio of the two effects to be different for the two stars.

We opt not to give in this paper error estimates of the derived RV amplitudes and the inferred companion masses. Although the formal error of the RV amplitude can be derived from the error on the beaming amplitude, the true error is much larger, as we have to include the error coming from the inaccuracy of determining the orbital phase. This requires a further analysis that we defer to the paper which presents our analysis of all {\it Kepler} lightcurves (Faigler \& Mazeh, in preparation). The companion mass error is even less known, as it depends not only of the amplitudes of the beaming and ellipsoidal effects, but also on the stellar primary mass, which is not well known at this stage. When we obtain spectra of the stars, we can better estimate the primary masses and give better constrains on the companion masses.

For both stars, the detection of the ellipsoidal and the beaming modulations was highly significant. The reflection effect, on the other hand, was less secure, as the amplitude of the detected modulations was only 5 and 3 times their respective formal errors (see the discussion in the previous paragraph). However, as the two main effects were significantly detected, we estimated these detections as being secure, and considered these stars as candidates for hosting a low-mass companion. In a separate paper (Faigler, Mazeh et al., in preparation) we present RV observations that confirm the existence of the two low-mass companions, and demonstrate good agreement between the BEER predicted period and velocity amplitude, and the RV ones. 

 The analysis of K8064 \& K6222 presented here was based on Q1 data only. After the BEER analysis 
{\it and} the performance of the RV observations that confirmed the photometric detection, the {\it Kepler} Q2 data was released. The Q2 lightcurves, with time span of additional 120 days, confirmed the detection of the photometric modulation. In particular, the newly derived photometric periods for both stars were consistent with the present photometric periods and the RV ones.  The whole photometric data and RV measurements will be presented in the coming paper (Faigler, Mazeh et al., in preparation). Here we present only the Q1 data and analysis, to show how we actually discovered the two stars, in order to demonstrate the potential of the BEER algorithm. 
 
\section{Discussion}        
\label{discussion}                            

We presented here a simple algorithm to detect candidates for low-mass {\it non-transiting} companions, using the {\it Kepler} and CoRoT lightcurves. The algorithm searches for the beaming effect, together with the ellipsoidal and the reflection modulations. 
The algorithm uses our prior knowledge of the stellar mass and radius, and the theory of tidal and beaming modulations, to verify that the ratio between the amplitudes of the beaming and ellipsoidal effects is as expected, and that the three effects have the correct relative phases. We expect the amplitudes of the effect to be on the order of 
$10$--$1000$ ppm.

At the level of precision needed for this work, 
stellar activity will contaminate the signal, and 
worse yet, it will do so at a timescale that is comparable to the signals 
of interest, i.e. the orbital period.
The associated flux modulations due to starspots can easily be larger 
than the expected beaming/ellipsoidal/reflection signal, and since the 
orbital period of the candidate is not known ahead of time, variations 
at the stellar rotation period can easily be of the correct duration 
to confound the BEER method.
Therefore the BEER method can find only {\it candidates}, 
and RV observations are absolutely required for any 
confirmation. This is similar to transiting searches, 
where RV follow-up measurements are crucial. 
To estimate statistically the yield of the BEER algorithm, 
the algorithm can be run on subsets of the data. RV follow up 
observations
of the candidates found in the subsets can help determine the false alarm rate 
as a function of spectral type and magnitude. 

The present version of the algorithm searches for a small-mass companion with a circular orbit. 
Obviously, an eccentric orbit will complicate the analysis, introducing higher harmonics of the orbital frequency. Radial-velocity modulation of eccentric orbit is a well understood effect, but the ellipsoidal modulation has still to be modeled carefully. Therefore, we assume in our analysis that the eccentricity contribution is small and defer more thorough analysis to the next stage of the development of the algorithm.

The two examples presented here demonstrated the potential of the {\it Kepler} lightcurves. The detection was done by using only the {\it Kepler} Q1 data, with timespan of only 33 days. We note that the detections were highly significant --- the derived amplitudes of the beaming effect in both stars were on the order of 100 ppm, while the formal errors on the modulation amplitudes were 2 and 5 ppm for K6222 and K8064, respectively (but see the discussion in the previous section). 
  If we ignore the stellar correlated noise \citep[for its implication see, for example,][]{pont06} we could expect the efficiency of the BEER algorithm to improve for longer lightcurves by up to $\sqrt{n}$, where $n$ is the number of observations. Therefore, we can expect to be able to detect a periodic modulation with an amplitude of a few ppm when we will have access to more {\it Kepler} data, at least for stars with the same noise level as that of K6222 and K8064. Thus, it might be possible in the near future to find candidates for planets with mass as small as $5$--$10\,M_\mathrm{Jup}$. \citet{mazeh10} showed that the CoRoT data is also accurate enough to detect the beaming and the ellipsoidal effects induced by a brown-dwarf companion. 

The long timespan of {\it Kepler} observations has one more advantage. If an analysis discovers an interesting candidate, follow-up RV measurements of this candidate can be obtained while the {\it Kepler} observations are still going on. This will enable the observer to compare not only the amplitude and period of the photometric beaming modulation with the RV observations, but also the phase of the detected beaming effect with the RV phase, confirming the existence of the low-mass companion with only very few RV observations.  Furthermore, if enough RV measurements were obtained and independent RV phase can be established, comparing the phase of the follow-up RV observations with that of the ellipsoidal modulation can give us some access to a possible lag between the two, which might indicate how the star is lagging after the tidal force exerted by its companion.

The attractiveness of the proposed approach is based on the fact that we have at hand on the order of a quarter of a million lightcurves (CoRoT and {\it Kepler} together) with  precision high enough to detect low-mass companions, depending on the stellar brightness. It is almost equivalent of having an RV survey of many thousands of stars with precision of $1$--$5$ km/s. This precision is enough to detect short-period low-mass-stellar, brown-dwarf companions, and even sometimes massive planets. 
 Our simulations indicate that in the {\it Kepler} Q1 data we can detect about 30\% of the brown-dwarfs with $40\,M_\mathrm{Jup}$ for stars with magnitude between $12$ and $13$ and stellar radii smaller than $3R_{\odot}$ (see Figure~2).
We might expect that when we have at hand three years of {\it Kepler} data, with a timespan of $\sim1000$ days instead of $\sim30$ day, our detection threshold will improve by $\sim\sqrt{1000/30}\simeq5$. 
This might be equivalent to have a modulation of $5 \times 40\,M_\mathrm{Jup}$ 
star in the present dataset. Consequently, we might expect that we will be able to detect 75\% of the brown-dwarf secondaries of that sample.
In upcoming papers (Faigler and Mazeh, in preparation) we list all the candidates found in the CoRoT and the {\it Kepler} data. These candidates, after confirmed, will increase the number of known low-mass-stellar/brown-dwarf companions and even massive planets.  

The approach presented here could have been conceived only because of the vision of \citet{loeb03} and \citet{zucker07}, who anticipated such detections to happen well before the launch of CoRoT and {\it Kepler}. We feel deeply indebted to the authors of these two papers and to the teams of the CoRoT and {\it Kepler} satellites, who built and are maintaining these missions, enabling us to search and analyze their unprecedented photometric data.


\section*{Acknowledgements}
We are indebted to Shay Zucker for helpful discussions and suggestions. We thank Dave Latham and Avi Shporer for careful reading of a previous version of the paper and illuminating comments.  
We thank to the anonymous referee whose comments and suggestions are highly appreciated. 

All the photometric data presented in this paper were obtained from the 
Multimission Archive at the Space Telescope Science Institute (MAST). 
STScI is operated by the Association of Universities for Research in 
Astronomy, Inc., under NASA contract NAS5-26555. Support for MAST for 
non-HST data is provided by the NASA Office of Space Science via grant 
NNX09AF08G and by other grants and contracts.

This research was supported by the ISRAEL SCIENCE FOUNDATION (grant No.
655/07).


{}

\label{lastpage}
\end{document}